\documentclass[10pt,showpacs,twocolumn,nofootinbib,aps,prd]{revtex4-1}
\usepackage{CJK}
\usepackage{amsmath,amssymb,bm,mathrsfs}
\usepackage{mathtools,mathptmx,array,slashed}
\usepackage{graphicx}
\usepackage{subfig}
\usepackage{ctable,multirow}
\usepackage{tabularx}
\usepackage[colorlinks=true,linkcolor=blue,citecolor=blue,urlcolor=blue]{hyperref}
\usepackage{dblfloatfix}

\def\be{\begin{equation}}
\def\ee{\end{equation}}
\def\bea{\begin{eqnarray}}
\def\eea{\end{eqnarray}}
\def\ba{\begin{aligned}}
\def\ea{\end{aligned}}

\begin{document}


\title{EHT-Constrained Analysis of Shadow Deformation in Quantum-Improved Rotating Non-Singular Magnetic Monopole}

\author{Gowtham Sidharth M.}
\email{gowthamsidharth.m2019@vitstudent.ac.in}
\address{School of Advanced Sciences Vellore Insitute of Technology, Melakottaiyur, Chennai, 600127, TamilNadu, India\\}

\author{Sanjit Das}
\email{sanjit.das@vit.ac.in}
\address{School of Advanced Sciences Vellore Insitute of Technology, Melakottaiyur, Chennai, 600127, TamilNadu, India\\}

\begin{abstract}
We studied the shadow cast by a rotating Bardeen black hole within the framework of asymptotically safe gravity. The null geodesics were analyzed using the Hamilton–Jacobi separation method to derive shadow observables. Our findings show that an increase in both the asymptotic safety parameter $(\omega)$ and the spin parameter $(a)$ leads to a decrease in the apparent shadow size and an increase in shadow distortion. The monopole charge $(g)$ of the black hole played an important role in the shadow profile. Furthermore, we compute the energy emission rate associated with varying values of the asymptotic safety parameter. 
\end{abstract}
\maketitle

\section{Introduction}

The General Theory of Relativity(GR), formulated by Einstein in 1915, changed our perception of gravity from a fundamental force to the curvature of space-time induced by mass and energy \cite{Einstein1916}. This framework accurately predicted planetary orbits, gravitational lensing, and gravitational waves detected a century later. GR also plays a crucial role in understanding the galactic dynamics and large-scale structure of the universe. Predicted purely by mathematics, black holes are the most fascinating objects in the known cosmos. They represent some of the simplest yet most enigmatic solutions to Einstein's equations\cite{Hawking1972}. Karl Schwarzschild derived one of the first solutions of the Einstein equations for a non-rotating mass \cite{Schwarzschild1916}, later extended by Reissner and Nordström to include electromagnetic fields \cite{Reissner,Nordström}.

For decades, theorists have debated the existence of black holes. Their plausibility gained traction when Oppenheimer and Snyder suggested that black holes could form through the gravitational collapse of massive stars \cite{Oppenheimer1939}. Observational discoveries have motivated the theoretical investigations into black hole solutions within GR. The Kerr solution is a pivotal advancement in describing rotating masses\cite{Kerr1963}, and was later generalized by Newman to include charge \cite{Newman}. Interestingly, these solutions show all the earlier black hole models as special cases.

While these classical solutions provide the geometric foundation of black hole spacetimes, recent attention has shifted toward observable signatures such as black hole shadows, which offer a direct probe of the underlying geometry. In particular, shadows of rotating black holes, including regular Bardeen-type spacetimes, have been extensively studied, and general frameworks for their construction in Newman–Janis–generated metrics are well established \cite{Tsukamoto2014, Tsukamoto2018}.

Within classical GR, the no-hair theorems imply that stationary astrophysical black holes are fully characterized by only three parameters\cite{Israel1967,Carter1971,Robinson1975}: mass, spin, and charge, and the limitation of GR lies in the validity of these solutions. If observational data reveal deviations from the Kerr description, we would come to terms with any of the following three conclusions: (1) the GR fails in the regime of strong gravity, (2) the compact objects we identify as black holes might have physical surfaces, or (3) naked singularities exist\cite{Cardoso2019,Shaikh2023}. Any of these findings would signal the need for new physics, challenging our fundamental understanding of gravity and the nature of spacetime.

Further limitations arise from the classical nature of black holes and general relativity, particularly at the intersection of quantum mechanics\cite{Birrell1982,Sagnotti}. The major issue is non-renormalizability. Unlike quantum field theories such as electrodynamics and the Standard Model, which can be systematically renormalized to remove infinities, GR becomes uncontrollable at high energy scales. Conventional Attempts to quantize gravity lead to divergences at high energies\cite{Sagnotti}.

As GR is consistent and effective, fundamental theory breaks down near singularities, where quantum gravitational effects are expected to dominate; no complete theory of quantum gravity currently exists to describe such regimes. Moreover, GR is not applicable in extreme environments such as the Planck scale, where quantum fluctuations are expected.

This inconsistency with quantum mechanics has motivated the search for quantum gravity. A major challenge in quantizing gravity is non-renormalizability~\cite {Birrell1982,Aharony1999,tHooft1974}. This remains a fundamental obstacle to the unification of gravity with the other fundamental forces.This renormalization problem has encouraged physicists to search for alternative systems, including asymptotic safety, as proposed by Steven Weinberg in the 1970s. Asymptotically Safe Gravity (ASG) postulates that gravity remains renormalizable if it exhibits an ultraviolet(UV) fixed point \cite{Weinberg1979,Reuter1998,Percacci2009,Niedermaier2006,Eichhorn2018,Percacci2017,Saueressig}.This UV point constrains the behavior of gravitational couplings at high energies, making quantum gravity consistent with large-scale general relativity near the Planck scale. This fixed point makes the gravity asymptotically safe. At low energies, the infrared limit is consistent with planetary motions. This makes the ASG a suitable candidate for the unified theory of gravity.~\cite{Litim2006,Falls2015,Donoghue,Bonanno2010,Narain2010,Litim2011,Falls2013,Shaposhnikov2010,Reuter2012,Hamber2009,
Percacci2003,Falls:2012}.In this work, we follow the structure of \cite{RuizTuiran2023} used for RG coupling of Reissner-Nordstrom spacetime.

This self-consistent quantum gravity framework presents a well-modified theory of gravity, which has profound implications for black hole physics, cosmology, and the structure of spacetime at high energies.Alternative approaches to quantum-corrected black hole geometries have also been explored, including models that modify causal structure and energy conditions in strong gravity regimes \cite{Battista2024}. Beyond theoretical consistency, black hole shadows created by event horizons against bright accretion material can now be tested with observational results. The Event Horizon Telescope image of the M87 black hole allows us to evaluate the Asymptotic Safety Gravity(ASG) predictions in comparison to General Relativity (GR). ~\cite{L1,Kocherlakota2021,Held2019,Cunha2018}.

Moreover, ASG suggests that quantum corrections near the event horizon can lead to observable modifications in shadow size and shape. Studying such anomalies would provide strong evidence for the quantum nature of gravity and support the ASG as a unified theory of gravity.~\cite{Brahma2021,Ashtekar2021,Barcelo2020,Platania2019,Gambini2020,Liu2020,Bardeen2018,Kumar2020,Alesci2019}.

While black hole shadows offer a macroscopic observational window, quantum effects such as Hawking radiation probe the microphysics near the horizon. In the vicinity of a black hole, the intense nature of spacetime curvatures triggers quantum fluctuations in the vacuum, resulting in the random birth and decay of short-lived virtual particle pairs. Typically, these pairs annihilate instantaneously, but near the event horizon, one particle can fall into the black hole while the other escapes because of quantum oscillations.

To a faraway observer, the escaping particles appear as thermal radiation; this was first proposed by Hawking, hence the radiation is termed Hawking radiation\cite{Hawkingb}. Over time, this energy loss causes the black hole to shrink slowly, changing its mass, temperature, and entropy. This slow evaporation continued until the black hole disappeared completely. The rate of energy emission depends on the temperature, which in turn is inversely proportional to its mass\cite{Hawkingb,Bekenstein1973}. Smaller black holes emit more radiation and evaporate faster than heavier black holes. The exact nature of the final stages of black holes is unknown. We studied the nature of the energy emission rate of a rotating Bardeen black hole in the context of ASG.

The Bardeen metric is chosen primaly because it represents a class of regular, singularity-free black holes, in contrast to traditional solutions such as Schwarzschild or Reissner-Nordström that feature central singularities where classical general relativity breaks down. This singularity-free nature aligns well with quantum gravity approaches such as asymptotically safe gravity (ASG), where running couplings are expected to remove singularities at the Planck scale. By focusing on the Bardeen metric, we can study gravitational effects precisely at the horizon scale, because its regular core ensures well-defined metric properties even near the center, enabling a detailed analysis of strong-field phenomena influenced by quantum corrections. The black hole shadow, an observable feature dependent on the spacetime geometry, serves as a powerful probe for detecting imprints of ASG. Furthermore, the Bardeen metric offers a convenient benchmark for comparison with other models because of its balance between physical realism and analytical tractability, making it well-suited for both analytical and numerical studies aimed at understanding how quantum gravitational modifications manifest around black holes and affect observable quantities.

Although the Bardeen metric is free of curvature singularities, its coupling to asymptotically safe gravity preserves this regularity\cite{ Platania2019,BonannoReuter2000}. In this work, we investigate the horizon-scale gravitational effects of asymptotic safety and examine their imprints on the black hole shadow. We begin by constructing the non-singular rotating Bardeen black hole within the ASG framework in Section 2. The geodesic equations are then derived using the Hamilton–Jacobi formalism in Section 3. In Section 4, we analyse the effective potential governing photon motion, followed by the computation and visualisation of black hole shadows in Section 5. The corresponding energy emission rate is studied in Section 6. Finally, in Section 7, we employ observational data to place constraints on the black hole shadow and the underlying model parameters.
\section{Non-Singular Blackhole in ASG} \label{BH}

A singularity-free black hole model was proposed by Bardeen, and many such models were later developed. The Bardeen black hole arises from a specific non-linear electrodynamics source~\cite{Bardeen1968,AyonBeatoGarcía2000} defined by the function $\mathcal{L}$ as follows: 

$$\mathcal{L}(F)=\frac{3}{2sg^2}\left(\frac{\sqrt{2g^2F}}{1+\sqrt{2g^2F}}\right)^{\frac{5}{2}}.$$

The Bardeen black hole's line element can be found as

\begin{equation}
    ds^2 = -f(r)dt^2+\frac{1}{f(r)}dr^2+r^2d\Omega^2,
\label{eq1}
\end{equation}
where $f(r)=1-\frac{2G_NMr^2}{(r^2+g^2)^{\frac{3}{2}}}$ in which $G_N$ is Newtonian Constant and $\textit{g}$ is the magnetic moment.

\subsection{Quantum Improved Bardeen Blackhole}
\label{S:2}

\subsubsection{The Running Newton Constant}
The effective average action $\Gamma_{k}$ is constructed by integrating out quantum fluctuations with momenta that are above the infrared cutoff scale $k$\cite{Reuter-1}. In the context of Quantum Einstein Gravity (QEG), this idea is implemented through the Euclidean path integral over metric configurations, starting from a specified action $S$.

To perform this procedure, a regulator function $R_{k}$ is introduced
so that modes with $p < k$ are suppressed. Thus, the construction provides a smooth Wilson’s renormalization group approach.

By changing the cutoff scale $k$, the family $\{\Gamma_{k}\}$ interpolates between theunderlying action and the full quantum effective action; in the ultraviolet limit,
$k \to \infty$, one recovers the bare action $S$, whereas in the infrared limit,$k \to 0$, the functional becomes the standard effective action $\Gamma$.

The dependence of $\Gamma_{k}$ on the running scale is governed by the Exact Renormalization Group Equation (ERGE), a functional differential equation that tracks the evolution of the action as the cutoff is lowered.

\begin{equation}
k\partial_{k}\Gamma_{k}=\frac{1}{2}Tr\left[k\partial_{k}R_{k}\left(\Gamma_{k}^{\left(2\right)}+R_{k}\right)^{-1}\right],  
\label{eq2}
\end{equation}

Here, $\Gamma^{(2)}_{k}$ represents the Hessian of $\Gamma_{k}$, which is formed by taking the second functional derivative with respect to the dynamical fields. Because obtaining an exact solution of (\ref{eq2}) is generally not feasible, one typical working method is to perform approximations. A widely used non-perturbative technique is to truncate the infinite-dimensional theory space and project the RG flow onto a finite subset of couplings. One selects an ansatz for $\Gamma_{k}$ that includes only a limited number of invariants and substitutes it into (\ref{eq2}). This procedure leads to a closed set of scale-dependent differential equations of the form:

\begin{equation}
k\partial_{k}g_{i}(k)=\beta_{i}(g_{1},g_{2},\cdots),
\label{eq3}
\end{equation}

Here, the functions $g_{i}(k)$ represent the dimensionless couplings corresponding to the individual invariants that are kept in the chosen truncation scheme. Here, we adopt the Einstein--Hilbert truncation \cite{Reuter1998, Saueressig}, which confines the renormalization group flow of $\Gamma_{k}$ to a two-parameter subspace characterized by the running Newton coupling $G_{k}$ and scale-dependent cosmological constant $\bar{\lambda}_{k}$. Under this approximation, the effective average action is modeled using the ansatz as follows:

\begin{equation}
\Gamma_{k}\left[g_{\mu\nu}\right]=\left(16\pi G_{k}\right)^{-1}\int
d^{d}x\sqrt{g}\left\{-R\left(g\right)+2\bar{\lambda}_{k}\right\},
\label{eq4}
\end{equation}
 We now extend the discussion to a general spacetime dimension $d$. When the Einstein--Hilbert truncation (\ref{eq4}) is inserted into the ERGE (\ref{eq2}), the flow equation reduces to a pair of coupled differential equations  of  dimensionless Newton coupling and the dimensionless cosmological constant. These quantities are introduced as follows:
\[
g_{k} \equiv k^{d-2} G_{k}, 
\qquad 
\lambda_{k} \equiv k^{-2} \bar{\lambda}_{k},
\]
as derived in \cite{Reuter1998, Saueressig}:

\begin{equation}
\partial_{t}g=\left(d-2+\eta _{N}\right)g , 
\label{eq5}
\end{equation}
and 
\begin{align}
\partial_{t}\lambda = {} & -\left(2 - \eta_{N}\right)\lambda \nonumber \\
& + \frac{1}{2} g (4\pi)^{\left(1-\frac{d}{2}\right)} \nonumber \\
& \times \Big[
2d(d+1)\Phi_{\frac{d}{2}}^{1}(-2\lambda)
- 8d\Phi_{\frac{d}{2}}^{1}(0) \nonumber \\
& \qquad - d(d+1)\eta_{N}\tilde{\Phi}_{\frac{d}{2}}^{1}(-2\lambda)
\Big].
\label{eq6}
\end{align}
Here, the parameter $t \equiv \ln k$ serves as the flow variable,which is often referred to as the
renormalization time. The quantity $\eta_{N}(g,\lambda)$ is introduced through the relation

\begin{equation}
\eta_{N}\left(g,\lambda\right)=\frac{gB_{1}\left(\lambda\right)}{1-gB_{2}\left(\lambda\right)}\;,  
\label{eq7}
\end{equation}
 \noindent and functions $B_{1}\left(\lambda \right)$ and $B_{2}\left(\lambda\right)$ are given by
\vspace{-0.1cm}
\begin{align}
B_{1}(\lambda) \equiv {} & \frac{1}{3} (4\pi)^{\left(1-\frac{d}{2}\right)} \nonumber \\
& \times \Big[
d(d+1)\Phi_{\frac{d}{2}-1}^{1}(-2\lambda)
- 6d(d-1)\Phi_{\frac{d}{2}}^{2}(-2\lambda) \nonumber \\
& \qquad - 4d\Phi_{\frac{d}{2}-1}^{1}(0)
- 24\Phi_{\frac{d}{2}}^{2}(0)
\Big], \nonumber \\
B_{2}(\lambda) \equiv {} & -\frac{1}{6} (4\pi)^{\left(1-\frac{d}{2}\right)} \nonumber \\
& \times \Big[
d(d+1)\tilde{\Phi}_{\frac{d}{2}-1}^{1}(-2\lambda)
- 6d(d-1)\tilde{\Phi}_{\frac{d}{2}}^{2}(-2\lambda)
\Big].
\label{eq8}
\end{align}
\noindent where the threshold functions for $p=1,2,...$ given by
\begin{eqnarray}
\Phi_{n}^{p}\left(s\right)&=\frac{1}{\Gamma\left(n\right)}\int_{0}^{\infty}dz\;z^{n-1}\frac{R^{\left(0\right)}\left(z\right)-zR^{\left(0\right)\prime}\left(z\right)}{\left[z+R^{\left(0\right)}\left(z\right)+s\right]^{p}}\;, 
\label{eq9}\\
\tilde{\Phi}_{n}^{p}\left(s\right)&=\frac{1}{\Gamma \left( n\right)}\int_{0}^{\infty }dz\;z^{n-1}\frac{R^{\left(0\right)}\left(z\right)}{\left[z+R^{\left(0\right)}\left(z\right)+s\right]^{p}}\;, 
\label{eq10} 
\end{eqnarray}
 depends on the cutoff function $R^{\left(0\right)}\left(z\right)$ with 
$z\equiv p^{2}/k^{2}$. $R^0$ is arbitrary, except for the two conditions $R^{\left(0\right)}\left(0\right) =1$ and $R^{\left(0\right)}\left(z\right)\rightarrow 0$ for $z\rightarrow\infty$. For explicit calculations we choose the exponential form:
\begin{equation}
R^{\left(0\right)}\left(z\right)=\frac{z}{e^{z}-1}\;.  
\label{eq11}
\end{equation}
From now on, we assume that $\bar{\lambda}$ $\ll k^{2}$ for our scales of interest, this means that $\lambda\left(k\right)\approx 0$ and we do not consider the influence of the running cosmological constant in the physics of spherically symmetric black holes. Consequently, the evolution of $g$ is  governed entirely by
\begin{equation}
k\partial_{k}g=\left(2+\eta_{N}\right)g=\beta\left(g\left(k\right)\right),  
\label{eq12}
\end{equation}
\noindent where the function $\eta_{N}\left(g\right)$ is given by
\begin{equation}
\eta_{N}\left(g\right)=\frac{gB_{1}}{1-gB_{2}}\;,  
\label{eq13}
\end{equation}
 with 
\begin{equation}
B_{1}\equiv B_{1}\left(0\right)=-\frac{1}{3\pi}\left[24\Phi_{2}^{2}\left(0\right)-\Phi_{1}^{1}\left(0\right)\right], 
\label{eq14}
\end{equation}
and
\begin{equation}
B_{2}\equiv B_{2}\left(0\right)=\frac{1}{6\pi}\left[18\tilde{\Phi}_{2}^{2}\left(0\right) -5\tilde{\Phi}_{1}^{1}\left(0\right)\right]. 
\label{eq15}
\end{equation}
 
Substituting function (\ref{eq11}) into definitions (\ref{eq9}) and (\ref{eq10}) leads to,
\begin{eqnarray*}
\Phi_{1}^{1}\left(0\right)&=\frac{\pi^{2}}{6},\qquad \Phi_{2}^{2}\left(0\right)=1 \\
\tilde{\Phi}_{1}^{1}\left(0\right)&=1,\qquad \; \; \, \tilde{\Phi}_{2}^{2}\left(0\right)=\frac{1}{2}
\end{eqnarray*}
and, 
\begin{equation}
B_{1}=\frac{\pi}{18}-\frac{8}{\pi},\; B_{2}=\frac{2}{3\pi}\,,
\label{eq16}
\end{equation}
with these expressions for $B_1$ and $B_2$, we substitute (\ref{eq13}) in (\ref{eq12}) to obtain the following expression for the $\beta$-function

\begin{equation}
\beta\left(g\right)=2g\left(\frac{1-\left(B_{2}-\frac{1}{2}B_{1}\right)g}{1-B_{2}g}\right),  
\label{eq17}
\end{equation}
with the following definitions
\begin{equation}
w\equiv-\frac{1}{2}B_{1},\; \omega^{\prime}=w+B_{2}\,, 
\label{eq18}
\end{equation}
 as a result, the $\beta$-function takes the form 
\begin{equation}
\beta\left(g\right)=2g\left(\frac{1-\omega^{\prime}g}{1-B_{2}g}\right),  
\label{eq19}
\end{equation}
\noindent where
\begin{equation}
w=\frac{4}{\pi}\left(1-\frac{\pi^{2}}{144}\right) ,\;\omega^{\prime}=\frac{14}{3\pi}-\frac{\pi}{36}\;.
\label{eq20}
\end{equation}
 The flow equation (\ref{eq12}) for $g(k)$, together with the beta function given in
(\ref{eq19}), implies the presence of two fixed points, denoted by $g_{\ast}$, which are identified by the condition $\beta(g_{\ast}) = 0$. One solution is the Gaussian fixed point $g_{\ast}^{\text{IR}} = 0$, which is attractive in the infrared. The second solution is a non-Gaussian fixed point that governs the ultraviolet behaviour of the theory:

\begin{equation*}
g_{\ast}^{\text{UV}}=\frac{1}{\omega^{\prime}}\;. \\
\end{equation*}
The ultraviolet (UV) fixed point acts as a boundary between two different coupling regimes: a weakly interacting domain with $g < g_{\ast}^{\mathrm{UV}}$ and a strongly interacting domain where $g > g_{\ast}^{\mathrm{UV}}$. Because the $\beta$-function (\ref{eq17}) is positive on the interval $g \in [0, g_{\ast}^{\mathrm{UV}}]$ and becomes negative beyond it, the RG flow of the dimensionless Newton coupling $g(k)$ naturally splits into three characteristic classes of trajectories \cite{BonannoReuter2000, Souma}:

\begin{enumerate}

\item Flows for which $g(k)$ remains negative at all scales. These run toward the infrared (IR) fixed point $g_{\ast}^{\mathrm{IR}} = 0$ as $k \rightarrow 0$.

\item Flows satisfying $g(k) > g_{\ast}^{\mathrm{UV}}$ for all $k$, which are driven into the UV fixed point $g_{\ast}^{\mathrm{UV}} = 1/\omega'$ in the limit $k \rightarrow \infty$.

\item Flows that evolve entirely within band $g(k) \in [0, g_{\ast}^{\mathrm{UV}}]$. These interpolate between two fixed points:

they approach $g_{\ast}^{\mathrm{IR}}$ as $k \rightarrow 0$ and tend towards $g_{\ast}^{\mathrm{UV}}$ for $k \rightarrow \infty$.

\end{enumerate}

The first group corresponds to an unphysical scenario because it requires a negative Newton constant, whereas the second group does not match the low-energy sector. Therefore, only the trajectories belonging to the third class were of physical relevance in the present analysis.

Differential equation (\ref{eq12}), together with the $\beta$-function (\ref{eq17}), 
admits an exact analytic solution \cite{BonannoReuter2000}, which reads
\begin{equation}
\frac{g}{\left(1-\omega^{\prime} g\right)^{\tfrac{w}{\omega^{\prime}}}}
=\frac{g(k_{0})}{\bigl[1-\omega^{\prime} g(k_{0})\bigr]^{\tfrac{w}{\omega^{\prime}}}}
\left(\frac{k}{k_{0}}\right)^{2}.
\label{eq21}
\end{equation}

In general, this expression cannot be inverted in closed form to solve directly for $g(k)$. However, it is also possible to obtain an accurate analytic approximation. The key observation is that the ratio $\tfrac{\omega^{\prime}}{w}$, computed from (\ref{eq20}), is numerically close to unity ($\tfrac{\omega^{\prime}}{w}\approx 1.18$). 
By setting $\tfrac{\omega^{\prime}}{w}=1$ in (\ref{eq21}), one arrives at a remarkably good approximation that retains all essential qualitative features of the RG flow \cite{BonannoReuter2000}:  
\begin{equation}
g(k)=\frac{g(k_{0})\,k^{2}}{w g(k_{0})\,k^{2}+\bigl[1-w g(k_{0})\bigr]\,k_{0}^{2}}.
\label{eq22}
\end{equation}

 The dimensionful Newton constant, defined by $G(k)\equiv g(k)/k^{2}$, then follows as
\begin{equation}
G(k)=\frac{G(k_{0})}{1+w\,G(k_{0})\,[k^{2}-k_{0}^{2}]}.
\label{eq23}
\end{equation}
Choosing $k_{0}=0$ as the reference scale, and identifying $G_{0}\equiv G(k_{0}=0)$ with 
the empirically measured Newton constant, this expression simplifies to
\begin{equation}
G(k)=\frac{G_{0}}{1+w\,G_{0}k^{2}}.
\label{eq24}
\end{equation}

 The constant $w$ depends explicitly on the choice of the cutoff function $R^{(0)}$ 
and is therefore non-universal. To ensure that $G(k)$ corresponds to a genuine observable, 
an additional source of non-universality must appear to compensate this scheme dependence. This is addressed in the following section.  

Despite the approximation $\tfrac{\omega^{\prime}}{w}\approx 1$, 
the running of $G(k)$ retains the correct qualitative structure.  
Expanding (\ref{eq24}) around the IR limit yeilds
\begin{equation}
G(k)=G_{0}-w G_{0}^{2}k^{2}+O(k^{4}), \qquad (k \to 0),
\label{eq25}
\end{equation}
such that $G(k)$ approaches the observed constant $G_{0}$ at low scales.  
In the opposite limit, for $k^{2}\gg G_{0}^{-1}$, we find that
\begin{equation}
G(k)\approx \frac{1}{w k^{2}}, \qquad (k \to \infty),
\label{eq26}
\end{equation}
This demonstrates that Newton’s constant vanishes asymptotically.  
This UV weakening of gravity is in line with expectations near the Planck scale 
and agrees with the results obtained in alternative approaches \cite{Polyakov}.


\subsubsection{Cutoff Identification}
\label{sec:math}
Drawing inspiration from the renormalization group treatment of the Uehling correction to the Coulomb potential in QED where the RG scale $k$ is linked to the inverse radial distance $r$ \cite{Ditt-Reuter, Uehling} we employ an analogous identification in the gravitational setting. In this approach, the running scale $k$ is tied directly to the geometric properties of the space-time background \cite{Reuter-1}:

\begin{equation}
k(P)=\frac{\gamma}{d(P)}\;,
\label{eq27}
\end{equation}
Here, $\xi$ is a dimensionless parameter that reflects our limited knowledge of the precise physical mechanism responsible for imposing the infrared cutoff. Function $d(P)$ denotes an invariant proper distance, serving as a geometric replacement for the coordinate-dependent quantity $r$, in line with the principles of general relativity. It represents the proper separation between a fixed reference point $P_{0}$ typically chosen as the origin and a simultaneous point $P$ in the space-time manifold. More explicitly, $d(P)$ is obtained by integrating the infinitesimal line element $\sqrt{|ds^{2}|}$ along a selected path $\mathcal{C}$, which is commonly considered to be a radial straight line in three-dimensional space \cite{BonannoReuter2000}:

\begin{equation}
d(P)=\int_{\mathcal{C}}\sqrt{|ds^{2}|}\;.
\label{eq28}
\end{equation}

\noindent Substituting (\ref{eq27}) into (\ref{eq24}) yields
\begin{equation}
G(P)=\frac{G_{0}\,d(P)^{2}}{d(P)^{2}+G_{0}\bar{w}}\;, 
\qquad \bar{w} \equiv w \gamma^{2},
\label{eq29}
\end{equation}
where  parameters $w$ and $\gamma$ are combined into a single effective constant $\bar{w}=w\gamma^{2}$. 

This new parameter cannot be derived solely from RG arguments; in principle, it must be fixed experimentally, for example through measurements of quantum corrections to the Newtonian potential \cite{Donoghue}. Importantly, $\bar{w}$ possesses properties that make it a natural switch for  quantum gravitational effect.
\begin{enumerate}
\item $\bar{w}$ is proportional to $\hbar$, reflecting its quantum origin.
\item $\bar{w}$ is the only constant in (\ref{eq29}) that governs the scale dependence of $G$.
\item In the classical limit $\bar{w}=0$, the standard Newton constant is recovered, $G(P)=G_{0}$.
\end{enumerate}

In general, the explicit form of $d(P)$ depends on the choice of integration path 
$\mathcal{C}$. However, for spherically symmetric space-times, the path can be Considered as a straight line from the origin to $P$, in which case $d(P)$ depends solely on the radial coordinate $r$ through the metric functions \cite{BonannoReuter2000,Tuiran}. Thus, we set $d(P)=d(r)$ and the cutoff identification simplifies to
\begin{equation}
k=\frac{\xi}{d(r)}\;.
\label{eq30}
\end{equation}
\noindent Substituting (\ref{eq30}) into (\ref{eq24}) yields the radial dependence of Newton constant:
\begin{equation}
G(r)=\frac{G_{0}\,d(r)^{2}}{d(r)^{2}+\bar{w}G_{0}}.
\label{eq31}
\end{equation}

For the Bardeen space-time  with fixed $(t,\theta,\phi)$, the proper 
distance $d(r)$ obtained from (\ref{eq28}) takes the form
\begin{equation}
d(r) = r + M\!\left(-\frac{r}{\sqrt{g^{2}+r^{2}}}+\operatorname{arcsinh}\!\left(\frac{r}{g}\right)\right) 
+ O(M^{2}).
\label{eq32}
\end{equation}

In practice, one is usually interested in a region sufficiently far from the central 
singularity ($r \gg 0$), where the method of RG-improvement of classical space-times is considered to be reliable. In this asymptotic regime, $d(r)$ approaches $r$ \cite{Tuiran}, thus,

\begin{equation}
d(r)\approx r, 
\qquad k\approx \frac{\gamma}{r}, 
\qquad r\gg 0.
\label{eq33}
\end{equation}
\noindent 

Using this approximation in (\ref{eq31}) leads to a simple expression for the 
scale-dependent Newton constant:
\begin{equation}
G(r)=\frac{G_{0}\,r^{2}}{r^{2}+\bar{w}G_{0}}\;.
\label{G4}
\end{equation}
\section{Rotating Bardeen black hole in asymptotically safe gravity}

In this section, with the Newman--Janis algorithm (NJA)\cite{Newman:1965}, we generalize the spherically symmetric Bardeen black hole solution in ASG to a Kerr-like rotating black hole solution.

The general static and spherically symmetric metric is
\begin{equation}
ds^2 = -f(r)\,dt^2 + g(r)^{-1}dr^2 + h(r)d\Omega^2,
\quad d\Omega^2 = d\theta^2 + \sin^2\theta d\phi^2.
\end{equation}

First, we transform to Eddington--Finkelstein coordinates:
\begin{equation}
du = dt - \frac{dr}{\sqrt{f(r)g(r)}}.
\end{equation}

Then the metric becomes
\begin{equation}
ds^2 = -f(r)du^2 - 2\sqrt{\frac{f(r)}{g(r)}}\,du\,dr + h(r)d\Omega^2.
\end{equation}

Using null tetrads, the contravariant metric reads
\begin{equation}
g^{\mu\nu} = -l^\mu n^\nu - l^\nu n^\mu + m^\mu \bar{m}^\nu + m^\nu \bar{m}^\mu.
\end{equation}

The key step of NJA is the complex transformation:
\begin{equation}
u \rightarrow u - ia\cos\theta, \quad
r \rightarrow r + ia\cos\theta.
\end{equation}

It is worth noting that the Newman--Janis algorithm involves an ambiguity in the complexification procedure of the radial coordinate and metric functions. Alternative approaches that avoid this complexification and provide a more systematic construction of rotating solutions have been proposed in the literature \cite{Azreg2014PRD}.

The metric functions transform as
\begin{equation}
f(r) \rightarrow F(r,\theta), \quad
g(r) \rightarrow G(r,\theta), \quad
h(r) \rightarrow \Sigma = r^2 + a^2\cos^2\theta.
\end{equation}

After applying the NJA and returning to Boyer--Lindquist coordinates, the rotating metric becomes
\begin{align}
ds^2 =& -\frac{(gh + a^2\cos^2\theta)\Sigma}{(k + a^2\cos^2\theta)^2}dt^2
+ \frac{\Sigma}{gh + a^2}dr^2 + \Sigma d\theta^2 \\
&- 2a\sin^2\theta \left[\frac{k - gh}{(k + a^2\cos^2\theta)^2}\right]\Sigma dt\,d\phi \\
&+ \Sigma \sin^2\theta \left[1 + \frac{a^2\sin^2\theta(2k - gh + a^2\cos^2\theta)}{(k + a^2\cos^2\theta)^2}\right] d\phi^2.
\end{align}

For a quantum corrected Bardeen black hole :
\begin{equation}
f(r) = g(r) = 1 - \frac{2Mr^2 G(r)}{(r^2 + g^2)^{3/2}}, \quad
h(r) = k(r) = r^2.
\end{equation}

Thus, the final rotating metric is
\begin{align}
ds^2 =& -\left(1 - \frac{2\rho r}{\Sigma}\right)dt^2
+ \frac{\Sigma}{\Delta_r}dr^2 + \Sigma d\theta^2
- \frac{4a\rho r \sin^2\theta}{\Sigma}dt\,d\phi \\
&+ \sin^2\theta \left(r^2 + a^2 + \frac{2a^2\rho r \sin^2\theta}{\Sigma}\right)d\phi^2,
\end{align}

where
\begin{equation}
2\rho = \frac{2Mr^3 G(r)}{(r^2 + g^2)^{3/2}} ,
\end{equation}
\begin{equation}
\Sigma = r^2 + a^2\cos^2\theta,
\end{equation}
\begin{equation}
\Delta_r = r^2 + a^2 - \frac{2Mr^4 G(r)}{(r^2 + g^2)^{3/2}} ,
\end{equation}
\begin{equation}
G(r)=\frac{G_{0}\,r^{2}}{r^{2}+\bar{w}G_{0}}\;.
\label{G4}
\end{equation}

The previous non-rotating variant can be obtained when a $\longrightarrow$ 0, where $a$ is the spin parameter that accounts for the blackhole spin. Setting a$\longrightarrow$0, g$\longrightarrow$0 and $\bar{w}$$\longrightarrow$0 gives schwarzchild blackhole.

The components of the Einstein tensor were computed using \textit{Mathematica} through explicit symbolic evaluation of the metric and its associated curvature tensors. In this analysis, we incorporate the effects of Asymptotic Safety by promoting the Newtonian gravitational coupling to a scale-dependent quantity, $G \to G(r)$. 

The resulting expressions demonstrate that the Einstein tensor depends explicitly on both $G(r)$ and its radial derivatives, reflecting the underlying renormalization group flow. In particular, the structure naturally organizes in terms of derivatives of the effective mass function $m(r) = G(r)\rho(r)$, highlighting the interplay between the running gravitational coupling and the matter distribution.
\begin{equation}
\begin{aligned}
G_{tt} &=
- \frac{a^2 r \sin^2\theta}{\Sigma^2}
( G(r)\rho'' + 2\rho' G'(r) + \rho G''(r) ) \\
&\quad + \frac{2\left[r^2(a^2 + r^2) - a^4 \sin^2\theta \cos^2\theta \right]}{\Sigma^3}
( G(r)\rho' + \rho G'(r) ) \\
&\quad - \frac{4 r^3 \rho G(r)}{\Sigma^3}
( G(r)\rho' + \rho G'(r) ) \\[1em]
G_{rr} &=
- \frac{2 r^2}{\Sigma \Delta}
( G(r)\rho' + \rho G'(r) ) \\[1em]
G_{\theta\theta} &=
- \frac{2 a^2 \cos^2\theta}{\Sigma}
( G(r)\rho' + \rho G'(r) ) \\
&\quad - r ( G(r)\rho'' + 2 G'(r)\rho' + \rho G''(r) ) \\[1em]
G_{\phi\phi} &=
- \frac{r \sin^2\theta (a^2 + r^2)^2}{\Sigma^2}
( G(r)\rho'' + 2 \rho' G'(r) + \rho G''(r) ) \\
&\quad - \frac{a^2 \sin^2\theta (r^2 + a^2)(a^2 + (2r^2 + a^2)\cos 2\theta)}{\Sigma^3}
( G(r)\rho' + \rho G'(r) ) \\
&\quad - \frac{4 a^2 r^3 \sin^4\theta}{\Sigma^3}
\rho G(r) ( G(r)\rho' + \rho G'(r) )\\[1em]
G_{t\phi} &=
\frac{a r (a^2 + r^2)\sin^2\theta}{\Sigma^2}
\left( G(r)\,\rho''(r) + 2 \rho'(r)\,G'(r) + \rho(r)\,G''(r) \right) \\
&\quad + \frac{a (a^2 + r^2)\left(a^2 \cos 2\theta + a^2 - 2r^2 \right)\sin^2\theta}{\Sigma^3}
\left( G(r)\,\rho'(r) + \rho(r)\,G'(r) \right) \\
&\quad + \frac{4 a r^3 \sin^2\theta}{\Sigma^3}
\, \rho(r)\,G(r)\left( G(r)\,\rho'(r) + \rho(r)\,G'(r) \right).
\label{GTT}
\end{aligned}
\end{equation}
\noindent
In order to compute the components of the energy--momentum tensor, following Refs.~\cite{Toshmatov2017,BenavidesGallego2020}, we adopt the standard orthonormal tetrad for the rotating Quantum corrected Bardeen black hole. The tetrad vectors are given by
\begin{equation}
\begin{aligned}
e^{\mu}_{(t)} &= \frac{1}{\sqrt{\Delta_r \Sigma}} \left( r^2 + a^2,\, 0,\, 0,\, a \right), \\
e^{\mu}_{(r)} &= \sqrt{\frac{\Delta_r}{\Sigma}} \left( 0,\, 1,\, 0,\, 0 \right), \\
e^{\mu}_{(\theta)} &= \frac{1}{\sqrt{\Sigma}} \left( 0,\, 0,\, 1,\, 0 \right), \\
e^{\mu}_{(\phi)} &= - \frac{1}{\sqrt{\Sigma \sin^2\theta}} \left( a \sin^2\theta,\, 0,\, 0,\, 1 \right).\label{OTT}
\end{aligned}
\end{equation}

\noindent
Combining Eqs.\ref{GTT} and \ref{OTT} with the Einstein field equations $G_{\mu\nu} = 8\pi T_{\mu\nu}$, the components of the energy--momentum tensor can be obtained.
\begin{align}
\mathcal{E} = -p_r &= \frac{r^2}{4\pi \Sigma^2}
\left( G(r)\,\rho'(r) + \rho(r)\,G'(r) \right),
\\[1em]
p_\theta = p_\phi &= \mathcal{E}
- \frac{1}{8\pi \Sigma}
\Big[
2\left( G(r)\,\rho'(r) + \rho(r)\,G'(r) \right) \\
&\quad + r \left( G(r)\,\rho''(r) + 2 \rho'(r)\,G'(r) + \rho(r)\,G''(r) \right)
\Big].
\end{align}

\noindent
The effective stress--energy tensor obtained in this framework exhibits the structure of an anisotropic fluid sourced by the combined effects of the running gravitational coupling $G(r)$ and the matter density profile $\rho(r)$. In particular, defining the effective quantity $\Phi(r) = G(r)\rho(r)$, the energy density and pressures depend only on the first and second derivatives of $\Phi(r)$.

A notable feature of this configuration is that the radial pressure satisfies $p_r = -\mathcal{E}$, indicating a vacuum-like equation of state in the radial direction. In contrast, the tangential pressures $p_\theta = p_\phi$ acquire additional contributions involving $\Phi''(r)$, leading to anisotropy in the system. This behavior is characteristic of regular black hole solutions, where deviations from classical vacuum arise due to effective matter sources.

From the perspective of asymptotic safety, the running of the Newton coupling $G(r)$ plays a crucial role in modifying the effective energy--momentum tensor. Even in the absence of exotic matter, the scale dependence of $G(r)$ induces nontrivial contributions to $\mathcal{E}$ and $p_i$, which can lead to violations of classical energy conditions in the deep interior region. Such violations are typically required to regularize the central singularity and are consistent with the expected quantum gravitational corrections.

At large distances, where $G(r) \to G_0$, the standard classical behavior is recovered, and the stress--energy tensor reduces to that of the underlying matter distribution $\rho(r)$. Therefore, the model smoothly interpolates between a quantum-corrected core and a classical asymptotic spacetime, providing a consistent description of a regular rotating black hole within the asymptotic safety framework.

\section{Weak Energy Condition for the ASG-Improved Bardeen Spacetime}

The weak energy condition (WEC) requires that the energy density measured by any timelike observer is non-negative\cite{Wald2010}. This is expressed as
\begin{equation}
T_{\mu\nu} u^\mu u^\nu \geq 0,
\end{equation}
for all timelike vectors $u^\mu$.

For an anisotropic fluid with stress-energy tensor of the form
\begin{equation}
T^\mu_{\ \nu} = \mathrm{diag}(-\mathcal{E},, p_r,, p_\theta,, p_\phi),
\end{equation}
the WEC reduces to the following set of inequalities:
\begin{align}
\mathcal{E} &\geq 0, \\
\mathcal{E} + p_r &\geq 0, \\
\mathcal{E} + p_\theta &\geq 0, \\
\mathcal{E} + p_\phi &\geq 0.
\end{align}
In axisymmetric spacetimes where $p_\theta = p_\phi$, the last two conditions are identical.

\subsection{Effective Stress-Energy Tensor}

For the asymptotically safe gravity (ASG) improved Bardeen spacetime, the metric function is constructed using a running Newton coupling $G(r)$ and a Bardeen-type mass profile $\rho(r)$. The geometry is characterized by
\begin{equation}
\Sigma = r^2 + a^2 \cos^2\theta.
\end{equation}

From the Einstein field equations, the effective energy density is given by
\begin{equation}
\mathcal{E} = \frac{r^2}{4\pi \Sigma^2} \left( G(r)\rho'(r) + \rho(r)G'(r) \right),
\end{equation}
while the radial pressure satisfies
\begin{equation}
p_r = -\mathcal{E}.
\end{equation}
The tangential pressures take the form
\begin{equation}
p_\theta = p_\phi = \mathcal{E} - \frac{1}{8\pi \Sigma}\left[2\left(G\rho' + \rho G'\right)* r\left(G\rho'' + 2\rho'G' + \rho G''\right)\right].
\end{equation}

\subsubsection{Radial Condition}

Using $p_r = -\mathcal{E}$, we immediately obtain
\begin{equation}
\mathcal{E} + p_r = 0.
\end{equation}
Thus, the radial component of the WEC is always satisfied.

\subsubsection{Energy Density Condition}

The condition $\mathcal{E} \geq 0$ requires
\begin{equation}
G(r)\rho'(r) + \rho(r)G'(r) \geq 0.
\end{equation}
This expression consists of two contributions:
\begin{itemize}
\item A classical term $G(r)\rho'(r)$ arising from the Bardeen profile,
\item A correction term $\rho(r)G'(r)$ induced by the running Newton coupling.
\end{itemize}

\subsubsection{Tangential Condition}

The non-trivial condition is
\begin{equation}
\mathcal{E} + p_\theta \geq 0,
\end{equation}
which yields
\begin{equation}
\mathcal{E} + p_\theta =
2\mathcal{E} - \frac{1}{8\pi \Sigma}
\left[
2(G\rho' + \rho G') + r(G\rho'' + 2\rho'G' + \rho G'')
\right].
\end{equation}

This expression depends on two key structures:
\begin{align}
G\rho' + \rho G', \
G\rho'' + 2\rho'G' + \rho G''.
\end{align}
The second term introduces higher-derivative corrections arising purely from the running of the Newton coupling.

At large radial distances ($r \to \infty$), the running coupling approaches a constant:
\begin{equation}
G(r) \to G_0, \quad G'(r), G''(r) \to 0.
\end{equation}
In this limit, the classical Bardeen solution is recovered and the WEC is satisfied.

In contrast, near the core ($r \to 0$), the running coupling behaves as
\begin{equation}
G(r) \sim r^2,
\end{equation}
and its derivatives contribute significantly. The higher-derivative terms in $\rho + p_\theta$ can dominate, leading to
\begin{equation}
\mathcal{E} + p_\theta < 0.
\end{equation}

The weak energy condition for the ASG-improved Bardeen spacetime satisfies the following properties:
\begin{itemize}
\item The radial condition $\mathcal{E} + p_r = 0$ is always satisfied.
\item The energy density remains non-negative for physically relevant parameter choices.
\item The tangential condition $\mathcal{E} + p_\theta \geq 0$ is violated in a small region near the black hole core.
\end{itemize}

Thus, we conclude that WEC is violated near the core but restored at large distances.
This violation arises from the running of the Newton coupling, which introduces effective higher-derivative corrections to the stress-energy tensor. Such localized violations are consistent with expectations from quantum gravity and are commonly associated with the regularization of spacetime singularities \cite{BambiModesto2013,NevesSaa2014}.

Fig 1((a) - (d)) and Fig 2((a) - (d)) shows the dependance of $\mathcal{E} $ and $\mathcal{E} + p_\theta$ on radius and angle of the quatum corrected bardeen blackhole. From the plots we can see that the energy density remains positive throughout the spacetime, thereby satisfying the first condition of the weak energy condition (WEC), namely $\mathcal{E}  \geq 0$. However, the quantity $\mathcal{E}  + p_{\theta}$, which determines both the null energy condition (NEC) and the angular component of the WEC, exhibits negative values in a localized region near the black hole core. This indicates a violation of the energy conditions arising from the quantum corrections encoded in the running Newton coupling.

Furthermore, the violation is not isotropic. As the rotation parameter $a$ increases, the region where $\mathcal{E}  + p_{\theta} < 0$ becomes increasingly localized and concentrated near the equatorial plane ($\theta = \pi/2$). This behavior can be attributed to the angular dependence of the metric function $\Sigma = r^2 + a^2 \cos^2\theta$, which attains its minimum value at the equator, thereby enhancing the effective stress-energy contributions in that region.

On the other hand, increasing the quantum correction parameter $\bar{w}$ tends to smooth the energy distribution and reduces the magnitude and extent of the violation. Consequently, the combined effect of rotation and quantum corrections leads to an anisotropic but controlled violation of the energy conditions, which plays a crucial role in the regularization of the black hole core.

\begin{figure}[htbp]
\centering

\subfloat[a=0.1; g=0.2; $\bar{w}=0.3$]{%
\includegraphics[width=0.45\textwidth]{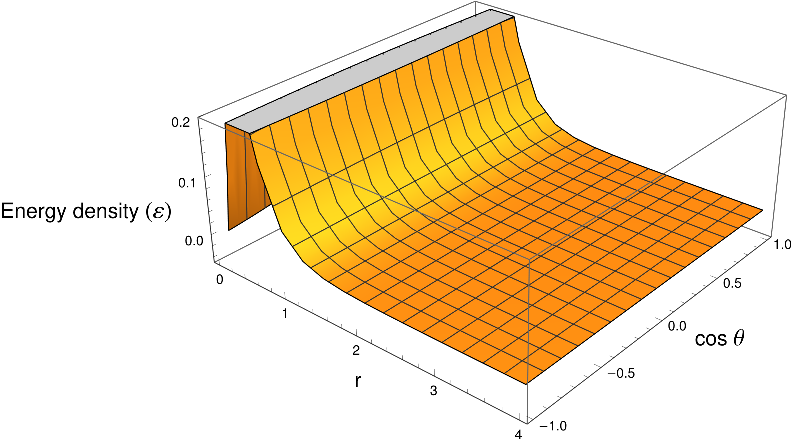}}
\hfill
\subfloat[a=0.9; g=0.2; $\bar{w}=0.3$]{%
\includegraphics[width=0.45\textwidth]{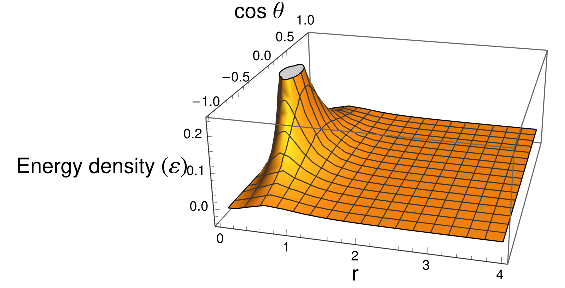}}

\vspace{0.4cm}

\subfloat[a=0.1; g=0.2; $\bar{w}=0.9$]{%
\includegraphics[width=0.45\textwidth]{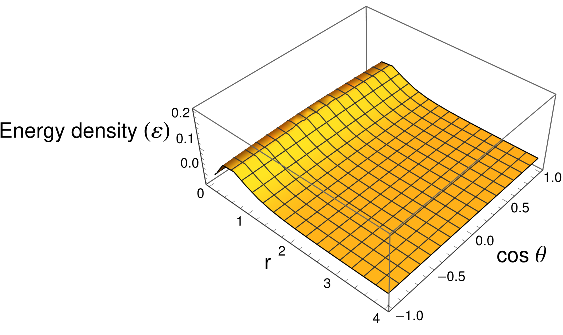}}
\hfill
\subfloat[a=0.9; g=0.2; $\bar{w}=0.9$]{%
\includegraphics[width=0.45\textwidth]{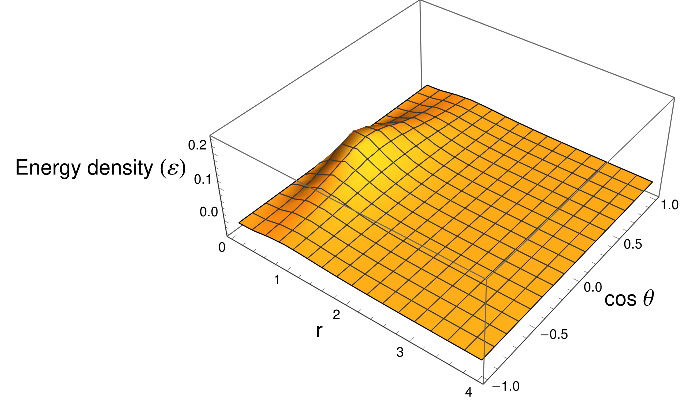}}

\caption{Dependence of energy density $(\mathcal{E})$ on radius and angle for a rotating quantum-corrected Bardeen black hole.}
\label{fig:wec_plots1}

\end{figure}

\begin{figure}[htbp]
\centering

\subfloat[a=0.1; g=0.2; $\bar{w}=0.3$]{%
\includegraphics[width=0.45\textwidth]{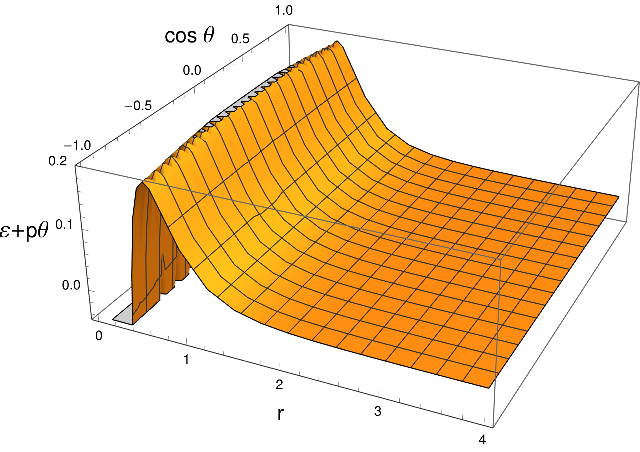}}
\hfill
\subfloat[a=0.9; g=0.2; $\bar{w}=0.3$]{%
\includegraphics[width=0.45\textwidth]{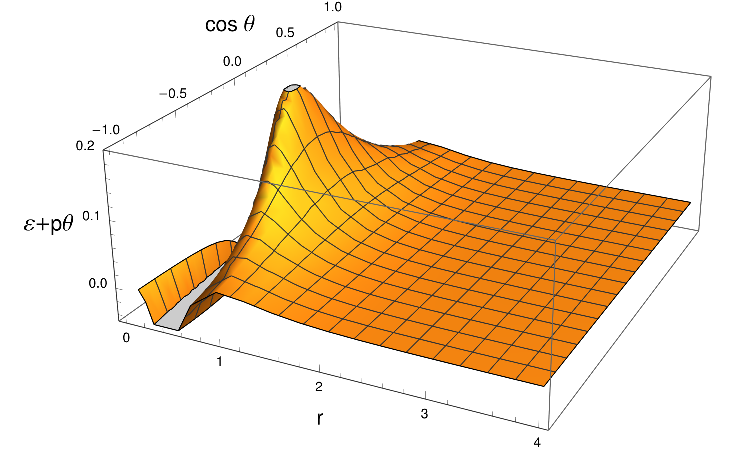}}

\vspace{0.4cm}

\subfloat[a=0.1; g=0.2; $\bar{w}=0.9$]{%
\includegraphics[width=0.45\textwidth]{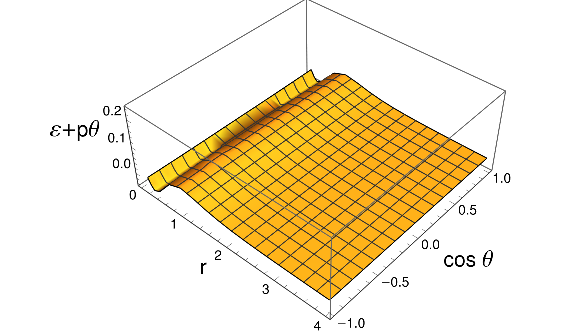}}
\hfill
\subfloat[a=0.9; g=0.2; $\bar{w}=0.9$]{%
\includegraphics[width=0.45\textwidth]{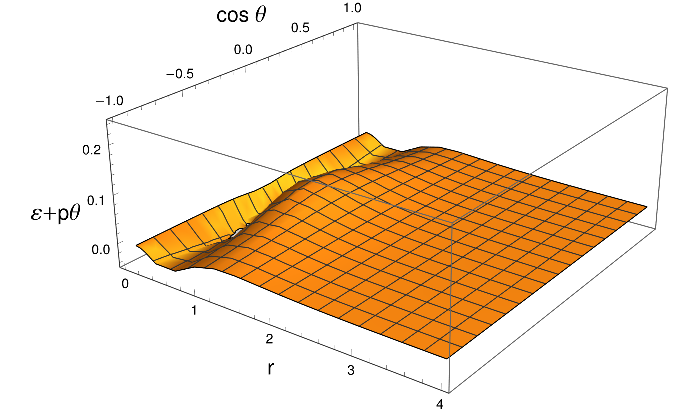}}

\caption{Dependence of $(\mathcal{E}+p_\theta)$ on radius and angle for a rotating quantum-corrected Bardeen black hole.}
\label{fig:wec_plots2}

\end{figure}

\section{Unstable photon orbits}

When a photon reaches the close vicinity of a black hole, it is deflected by a strong deflection and passes around the equator at least once before reaching the distant observer. Photons caught in this unstable orbit were responsible for the photon ring around the black hole. The variable separation Hamilton-Jacobi method is used to construct the equations for photon orbit. 

The general version of the Hamilton-Jacobi equation is
\begin{equation}\label{h1}
\frac{\partial S}{\partial \lambda} = -\frac{1}{2}g^{\mu\nu}\frac{\partial S}{\partial x^{\mu}} \frac{\partial S}{\partial x^{\nu}},
\end{equation}
where the affine parameter to the Jacobi action $S$ is  $\lambda$ and it corresponds to
\begin{equation}\label{h2}
S = \frac{1}{2}m^2 \lambda - E t + S_r(r)+S_{\theta}(\theta)+ L \phi ,
\end{equation}
with $m$ being the rest mass, the constants of motion, $E$ and $L$ are the corresponding energies and angular momentum . 

Since photons don't have rest masses, the previous equation changes to
\begin{equation}\label{h2}
S =  - E t + L \phi + S_r(r)+S_{\theta}(\theta).
\end{equation}

Following~\cite{Chandrasekhar1983} the solution for $S_r$ and $S_\theta$ yeilds
\begin{equation}
\Delta_r\left (\frac{\partial S_r}{\partial r}\right)^2 = (a L -(a^2+r^2)E)^2-\Delta_r((a E-L)^2+\mathcal{K}) = \mathcal{R}(r),
\end{equation}

\begin{equation}
\frac{\partial S_r}{\partial r} = \frac{\sqrt{\mathcal{R}(r)}}{\Delta_r},
\end{equation}

\begin{equation}
\left(\frac{\partial S_{\theta}}{\partial \theta}\right)^2 = \mathcal{K} - \left(\frac{L^2}{\sin^2 \theta}-a^2E^2\right)\cos^2 \theta = \Theta(\theta),
\end{equation}

\begin{equation}
\frac{\partial S_{\theta}}{\partial \theta} =\sqrt{\Theta(\theta)},
\end{equation}
where $\mathcal{K}$ is the sepaeration constant.

The equation of geodesic motion can be expressed as~\cite{Hou2018}
\begin{equation}\label{s1}
\Sigma \dot{t} = \frac{((r^2+a^2)E-aL)(r^2+a^2)}{\Delta_r}-a(aE \sin^2\theta-L),
\end{equation}

\begin{equation}
\Sigma\dot{r} =\sqrt{\mathcal{R}(r)},
\end{equation}

\begin{equation}
\Sigma \dot{\theta} =\sqrt{\Theta(\theta)},
\end{equation}

\begin{equation}\label{s2}
\Sigma \dot{\phi} = \frac{a((r^2+a^2)E-aL)}{\Delta_r}-\frac{aE \sin^2 \theta-L}{\sin^2 \theta},
\end{equation}
where
\begin{equation}
\mathcal{R}(r)= (a L -(a^2+r^2)E)^2-((a E-L)^2+\mathcal{K})\Delta,
\end{equation}

\begin{equation}
\frac{\partial S_r}{\partial r} = \frac{\sqrt{\mathcal{R}(r)}}{\Delta_r},
\end{equation}

\begin{equation}
\Theta(\theta)= \mathcal{K} - \left(\frac{L^2}{\sin^2\theta}-a^2E^2\right)\cos^2\theta,
\end{equation}

Consequently, the photon trajectory was calculated using the following two impact parameters
\begin{equation}
\xi =\frac{L}{E} ,\eta = \frac{\mathcal{K}}{E^2}.
\end{equation}

In terms of impact parameters (18) and (20) becomes,
\begin{equation}\label{r1}
\mathcal{R}_p(r) = (a \xi -(a^2+r^2))^2-((a-\xi)^2+\eta )\Delta,
\end{equation}

\begin{equation}
\Theta_p = \eta -\left(\frac{\xi^2}{\sin^2\theta} - a^2\right)\cos^2\theta.
\end{equation}

\subsection{Shadows}

The shadow boundry is determined by spherical photon orbits, which are null geodesics that remain at a constant radial coordinate. These orbits satisfy the conditions
\begin{equation}\label{r2}
\mathcal{R}_p(r)=0, \qquad \frac{d\mathcal{R}_p(r)}{dr}=0.
\end{equation}
The additional condition
\begin{equation}
\frac{d^2 \mathcal{R}_p(r)}{dr^2} < 0
\end{equation}
indicates that the orbit is radially unstable, which is the typical
behavior of spherical photon orbits that determine the boundary of the
black hole shadow.

Solving (\ref{r1}) and (\ref{r2}), $\xi$ and $\eta$ can be written as
\begin{equation}
\xi = \frac{- 4 r \Delta_r + a^2\Delta'_r + r^2 \Delta'_r}{a \Delta'_r},
\end{equation}
\begin{equation}
\eta = \frac{r^2(16 a^2\Delta_r-16\Delta^2_r+8r\Delta_r\Delta'_r-r^2\Delta'^2_r)}{a^2 \Delta'^2_r}.
\end{equation}

It is to note that the celestial coordinates ( $\alpha$ and $\beta$)  determine the black hole's shape. This celestial coordinate can be plotted to provide the shadow image. 
\begin{equation}
\alpha = \lim_{r_0 \rightarrow \infty}(-r^2_0 \sin\theta \frac{d \phi}{d r}|\theta \rightarrow i),
\end{equation}

\begin{equation}
\beta = \lim_{r_0 \rightarrow \infty}(-r^2_0 \frac{d \theta}{d r}|\theta \rightarrow i).
\end{equation}

The observer point is expressed as a function of the impact parameter $\xi$ and $\eta$ using geodesic equations (\ref{s1}) -- (\ref{s2}),
\begin{equation}
\alpha =- \frac{\xi}{\sin( i)},
\end{equation}

\begin{equation}
\beta = \pm \sqrt{\eta+a^2 \cos^2(i) -\xi^2 \cot^2(i)}.
\end{equation}

In the equatorial plane, the above equations are reduced as
\begin{equation}
\alpha = -\xi,
\end{equation}

\begin{equation}
\beta =\pm \sqrt{\eta}.
\end{equation}

\begin{figure*}[!htp]
\centering
\includegraphics[scale=1.5]{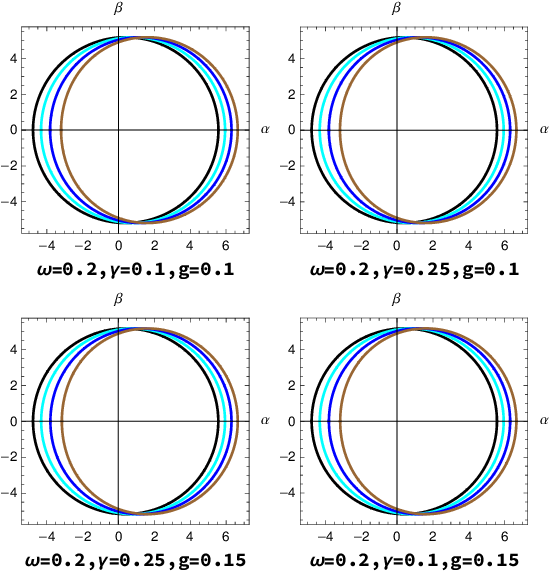}
\caption{Shadow plots of rotating Bardeen ASG black hole for different coupling variable $\omega$ and $\gamma$ with varying spin parameter $a$(Black=0.2;Cyan=0.4;Blue=0.6;Brown=0.8) values for constant g as $g$=0.25.}
\label{Fig1}
\end{figure*}

\begin{figure*}[!htp]
\centering
\includegraphics[scale=1.0]{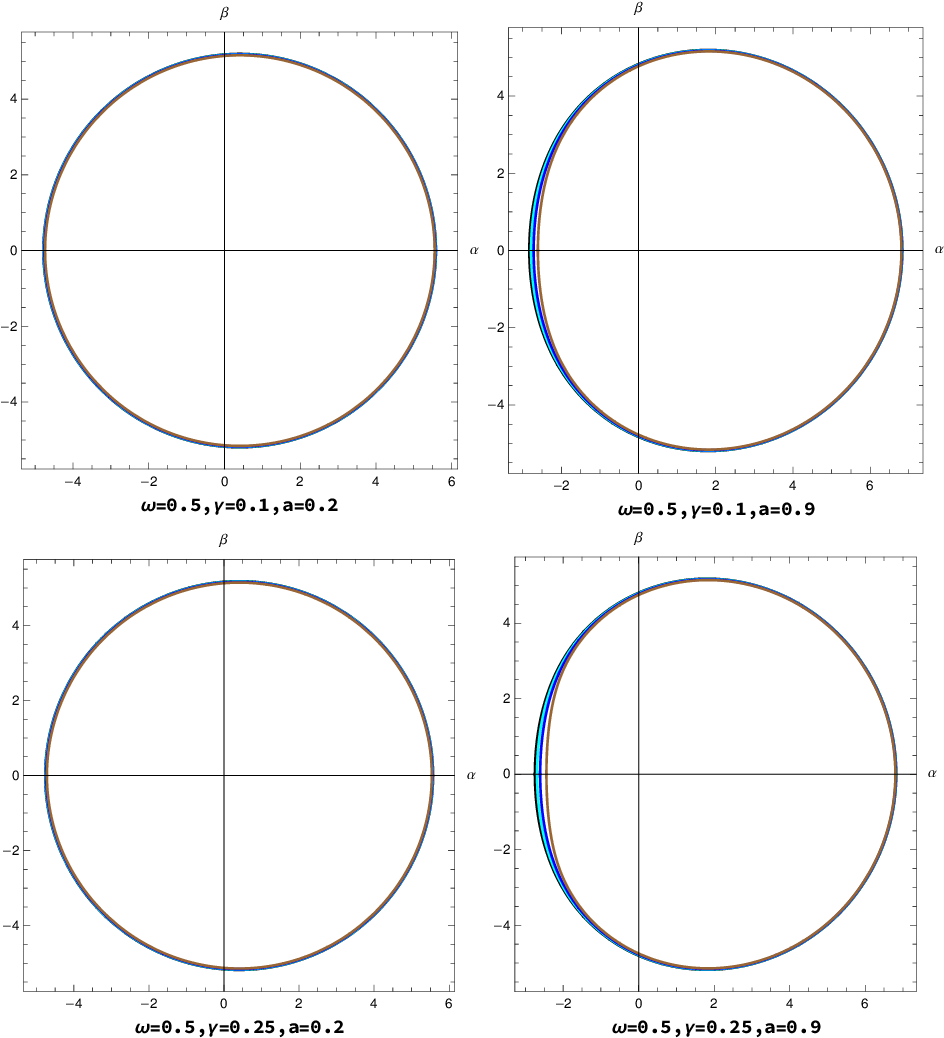}
\caption{Shadow plots of rotating Bardeen ASG black hole for different g values((Black=0.0;Cyan=0.1;Blue=0.15;Brown=0.2)) with varying spin parameter $a$ and $\gamma$values for constant  $\omega = 0.5$. }
\label{Fig2}
\end{figure*}

A distorted shadow image for a rotating blackhole owing to the frame dragging effect is expected. In Fig.\ref{Fig1} we can  see a increase in distortion for higher spin and $\gamma$ values.  Fig.\ref{Fig2} shows a slight increase in the distortion for higher $\omega$ and \textit{g}  value.


\section{Energy emission rate}

In this section, we compute the energy emission rate of a rotating Bardeen black hole in asymptotically safe gravity. The expression for the energy emission rate is given as
\begin{equation}
\frac{d^2 E(\Omega)}{d\Omega dt} = \frac{2\pi^2 \sigma_{lim}}{e^{\Omega/T}-1}\Omega^3,
\end{equation}
where $\Omega$ is the frequency of photon and the limiting constant $\sigma_{lim}$ is
\begin{equation}
\sigma_{lim} \approx \pi R^2_s
\end{equation}
and Hawking temperature $T$ is defined as
\begin{equation}
T =\lim_{\theta=0,r\to r_+} \frac{\partial_r \sqrt{-g_{tt}}}{2 \pi \sqrt{g_{rr}}},
\end{equation}
where $r_+$ is the outer event horizon.


\begin{figure*}[!htp]
\centering
\includegraphics[scale =0.6]{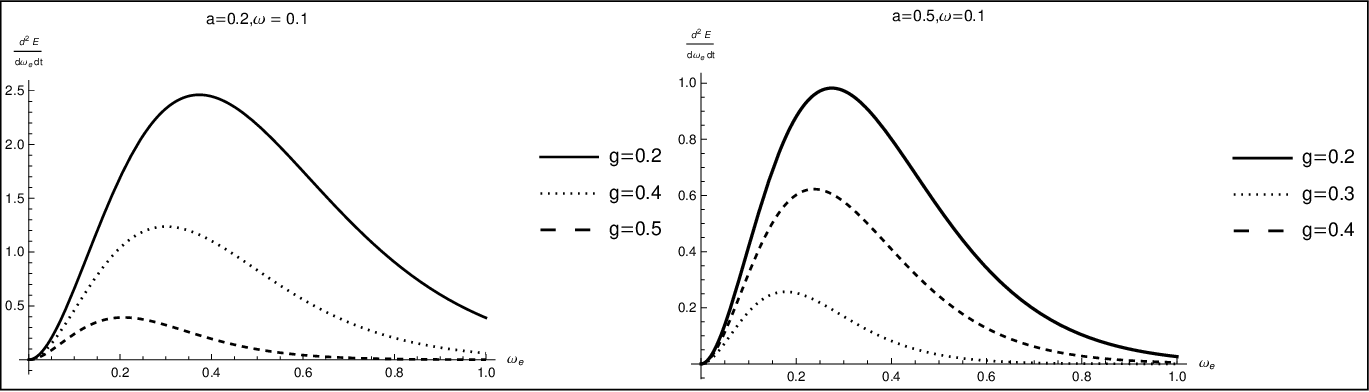}
\caption{Energy emission rate for $\omega$ =0.1  and different values of $a$ and $g$}
\label{Fig3}
\end{figure*}

The energy emission rate with respect to the frequency of photons is plotted in Fig. \ref{Fig3}. The emission rate increased with an increase in the monopole charge, and for higher spin values,  there was been a considerable decrease in the emission rate.

\section{OBSERVATIONAL CONSTRAINT ON ROTATING BLACK HOLES IN ASG}

Observational data from the blackholes M87* and Sgr A*  can be used to predict the possibility of a Bardeen blackhole in ASG. To define the observational quantites lets construct a shadow profile that is symmetric on the $\alpha$ -  axis ($\beta =0$). The centeroid of the shadow image with  area element $\mathcal{A}$ is defined as;

\begin{equation}
\alpha_c =\frac{\int \alpha d\mathcal{A}}{\int d\mathcal{A}},  \beta_c =0.
\end{equation}

The radial distance $l(\phi)$ from the geometric center of the shadow image to a point on its boundary, measured at an angle 
$\phi$ relative to the $\alpha-axis$, is given by:

\begin{equation}
\ell(\phi) ^2= (\alpha(\phi) - \alpha_c)^2 + \beta(\phi)^2
\end{equation}
and the average radius given by,

\begin{equation}
R^2_{avg} =\frac{1}{2 \pi} \int^{2\pi}_0 \ell^2(\phi) d\phi
\end{equation}
Using average radius we get the deviation from circularity \cite{Bambi},
\begin{equation}
\Delta C = \frac{1}{R_{\text{avg}}} \sqrt{\frac{1}{2\pi} \int_0^{2\pi} \left(\ell(\phi) - R_{\text{avg}}\right)^2 \, d\phi},
\end{equation}

Expressing $R_{avg}$ and $\Delta C$in $r_{ph}$ instead of $\phi$ for conviencce,
\begin{equation}
R_{\text{avg}}^2 = \frac{1}{\pi} \int_{r_{\text{ph}-}}^{r_{\text{ph}+}} \left(\beta'(\alpha - \alpha_c) - \beta \alpha'\right) dr_{\text{ph}},
\end{equation}
\begin{equation}
\Delta C = \frac{1}{R_{\text{avg}}} \sqrt{\frac{1}{\pi} \int_{r_{\text{ph}-}}^{r_{\text{ph}+}} \left(\beta'(\alpha - \alpha_c) - \beta \alpha'\right) \left(1 - \frac{R_{\text{avg}}}{\ell}\right)^2 dr_{\text{ph}}},
\end{equation}
and the geometric center,
\begin{equation}
\alpha_c = \frac{\int_{r_{\text{ph}-}}^{r_{\text{ph}+}} \alpha \beta \alpha' \, dr_{\text{ph}}}{\int_{r_{\text{ph}-}}^{r_{\text{ph}+}} \beta \alpha' \, dr_{\text{ph}}}, \beta =0.
\end{equation}

Here, $r_{ph}$ are at values where the shadow boundary passes through the $\alpha$- axis; hence roots $\beta(r_{ph}) = 0$ gives the $r_{ph+}$ and $r_{ph-}$.

The observational data for Sgr A\(^*\) does not currently provide information on the deviation parameter \(\Delta C\). However, for M87\(^*\), the data imposes a constraint of \(\Delta C \lesssim 0.1\) for an inclination angle of \(17^\circ\) \cite{L1,L5,L6,L12,L17}. This inclination angle corresponds to the orientation of the relativistic jets associated with M87\(^*\).

The fractional deviation of the shadow diameter $\delta$ have been used in EHT papers,The fractional deviation parameter $\delta$  is the ratio of shadow diameter to that of a Schwarzschild blackhole,
\begin{equation}
\delta=\frac{d_{sh}}{d_{sh,sch}}-1=\frac{R_{sh}}{3\sqrt{3}M} -1
\end{equation}

The EHT collaboration examined the shadow of Sgr A* using two independent datasets for its mass and distance, derived from VLTI and Keck observations, and imposed a constraint on  parameter \( \delta \) as \cite{L12,L17},

\begin{equation}
\delta = \begin{cases}-0.08^{+0.09}_{-0.09} & (VLTI)\\  -0.04^{+0.09}_{-0.10} & (Keck) \end{cases}
\end{equation}

Thus, the observational constraints from the VLTI and Keck data restrict the fractional deviation parameter to the range \( -0.14 < \delta < 0.01 \). Observational data indicates that values above \(\theta_0 = 50^\circ\) are disfavored. 

\begin{figure*}
     \centering
     \begin{minipage}{0.4\textwidth}
	\includegraphics[scale =0.8]{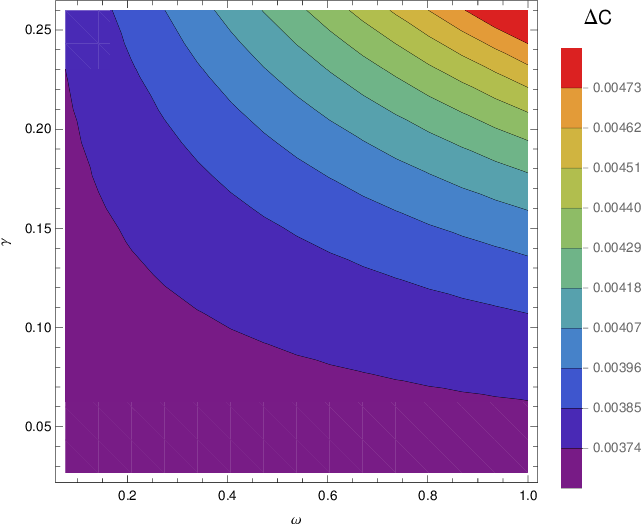}
	\end{minipage}
     \hfill
	\begin{minipage}{0.35\textwidth}
	\includegraphics[scale =0.8]{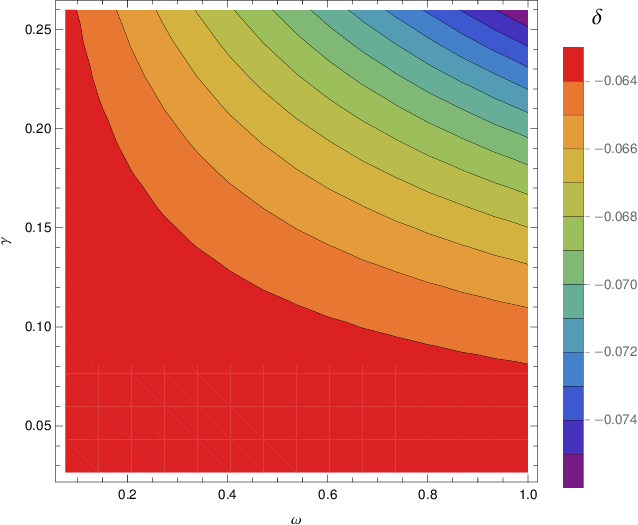}
	\end{minipage}
\caption{The contours corresponding to varying values of \( \Delta c \) (left)  and \(\delta \) (right) are 	depicted in the \( \omega \) vs \( \gamma \)  parameter space for fixed parameters \( a = 0.9 \) and \( g = 0.20 \) for the incilantion angle $17^o$.}
\label{Fig4}
\end{figure*}

\begin{figure*}
     \centering
     \begin{minipage}{0.4\textwidth}
	\includegraphics[scale =0.8]{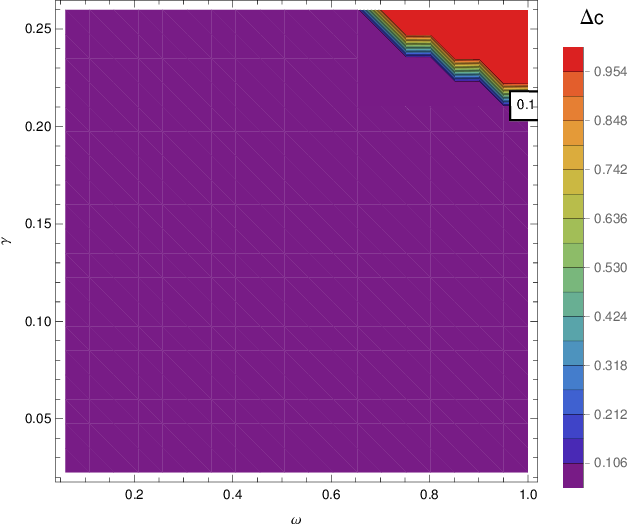}
	\end{minipage}
     \hfill
	\begin{minipage}{0.35\textwidth}
	\includegraphics[scale =0.8]{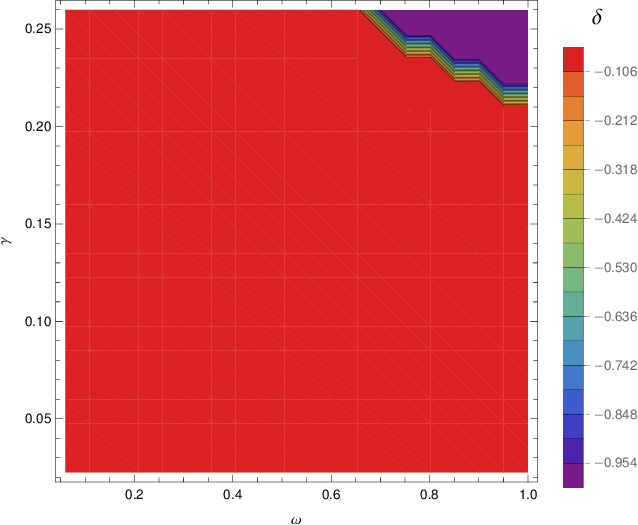}
	\end{minipage}
\caption{The contours corresponding to varying values of \( \Delta c \) (left)  and \(\delta \) (right) are 	depicted in the\( \omega \) vs \( \gamma \)  parameter space for fixed parameters \( a = 0.9 \) and \( g = 0.25 \)  for the incilantion angle $90^0$.}
\label{Fig5}
\end{figure*}

\begin{figure*}
     \centering
     \begin{minipage}{0.3\textwidth}
	\includegraphics[scale =0.8]{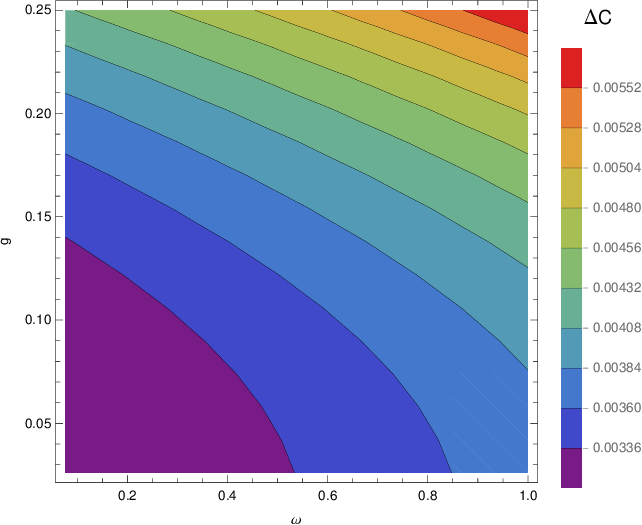}
	\end{minipage}
     \hfill
	\begin{minipage}{0.35\textwidth}
	\includegraphics[scale =0.8]{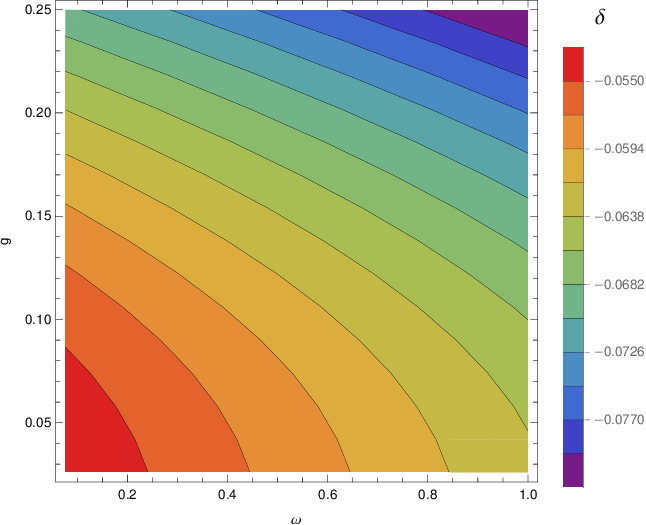}
	\end{minipage}
\caption{The contours corresponding to varying values of \( \Delta c \) (left) and \(\delta \) (right)  are 	depicted in the \( \omega \) vs g  parameter space for fixed parameters \( a = 0.9 \) and \( \gamma = 0.25 \) for the incilantion angle $17^0$.}
\label{Fig6}
\end{figure*}

\begin{figure*}
     \centering
     \begin{minipage}{0.4\textwidth}
	\includegraphics[scale =0.8]{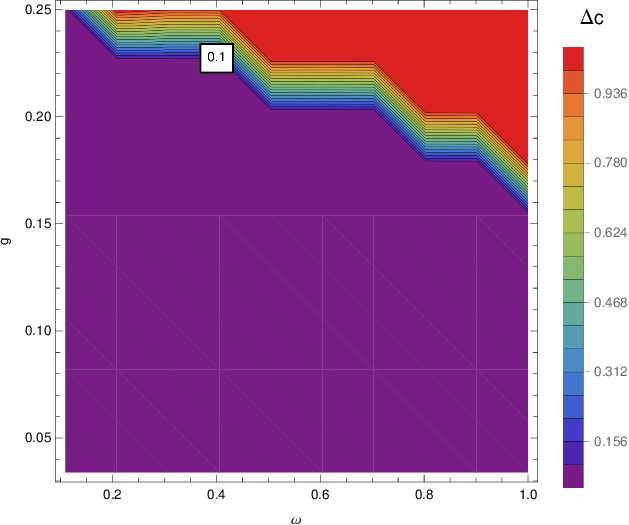}
	\end{minipage}
     \hfill
	\begin{minipage}{0.35\textwidth}
	\includegraphics[scale =0.8]{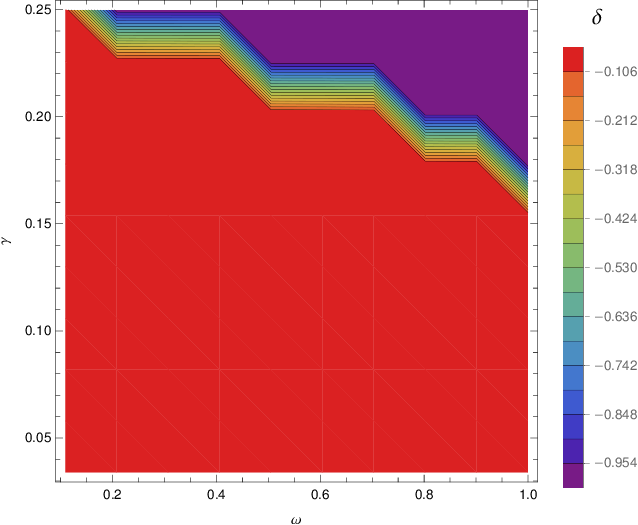}
	\end{minipage}
\caption{The contours corresponding to varying values of \( \Delta c \)   (left) and \(\delta \) (right) are 	depicted in the \( \omega \) vs g parameter space for fixed parameters \( a = 0.9 \) and \( \gamma = 0.25 \) for the incilantion angle $90^0$.}
\label{Fig7}
\end{figure*}

The rotating Bardeen black hole in the context of asymptotically safe gravity (ASG) is characterized by a set of parameters: mass (\(M\)), spin (\(a\)), magnetic monopole charge(\(g\)), and two coupling parameters (\(\gamma\) and \(\omega\)) that emerge from the ASG framework. In this study, we aim to constrain these parameters using observational data from the supermassive black holes M87\(^*\) and Sgr A\(^*\). We mainly focus on the coupling parameters, \(\gamma\) and \(\omega\), and their relationship to the observational constraints.

we estabish two parameter spaces,one which consist of $\gamma \ and \ \omega$(Fig. \ref{Fig4} and Fig. \ref{Fig5}) keeping a and g fixed and the other consits of $\omega \ and \ g$ (Fig. \ref{Fig6} and Fig. \ref{Fig7})to find all plausible values that would satisfy the  \(\Delta C \lesssim 0.1\) and also fractional deviation parameter.Given that \(\Delta C\) is known to increase with inclination angle, we extended our analysis to \(\theta_0 = 90^\circ\) instead of \(\theta_0 = 17^\circ\) to identify the maximum allowable values of \(\Delta C\).Based on the contours of parameter space,we conclude that the $\gamma$ values are constrained to be less than 0.25 and g value to be less than 0.15  .

\begin{figure*}
     \centering
     \begin{minipage}{0.4\textwidth}
	\includegraphics[scale =0.8]{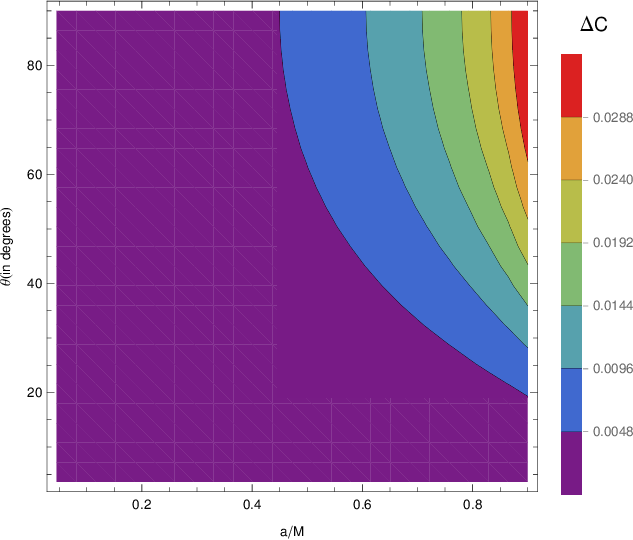}
	\end{minipage}
     \hfill
	\begin{minipage}{0.35\textwidth}
	\includegraphics[scale =0.8]{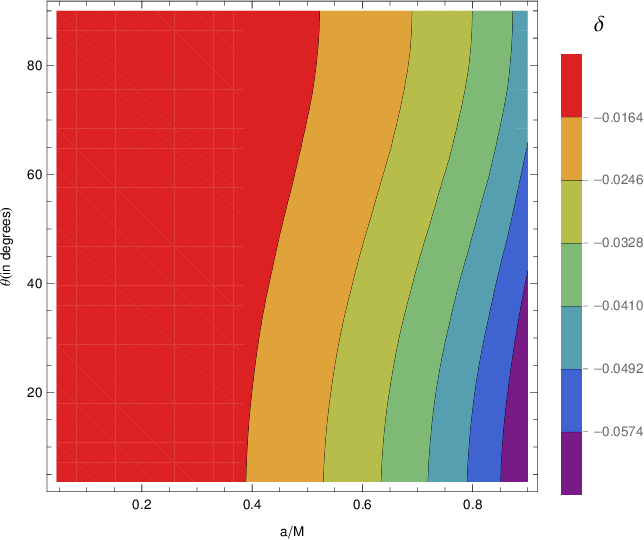}
	\end{minipage}
\caption{The contours corresponding to varying values of \( \Delta c \)  (left) and \(\delta \) (right)   are 	depicted in the a vs. \( \theta \) parameter space for fixed parameters \( \omega = 0.5, \gamma = 0.25 \) and \( g = 0.15 \).}
\label{Fig8}
\end{figure*}

\begin{figure*}
     \centering
     \begin{minipage}{0.4\textwidth}
	\includegraphics[scale =0.8]{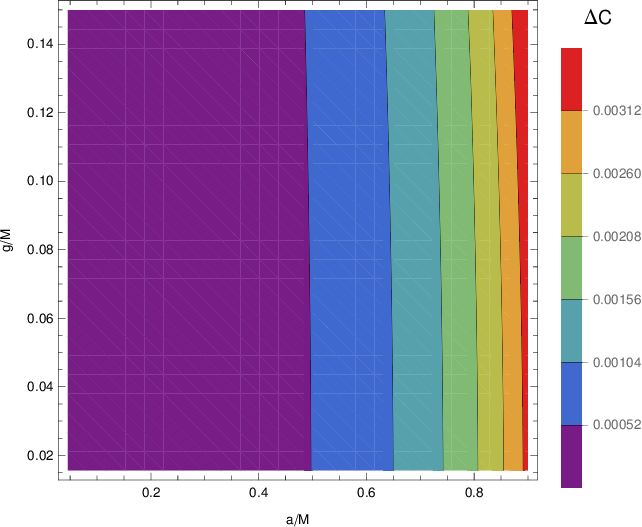}
	\end{minipage}
     \hfill
	\begin{minipage}{0.35\textwidth}
	\includegraphics[scale =0.8]{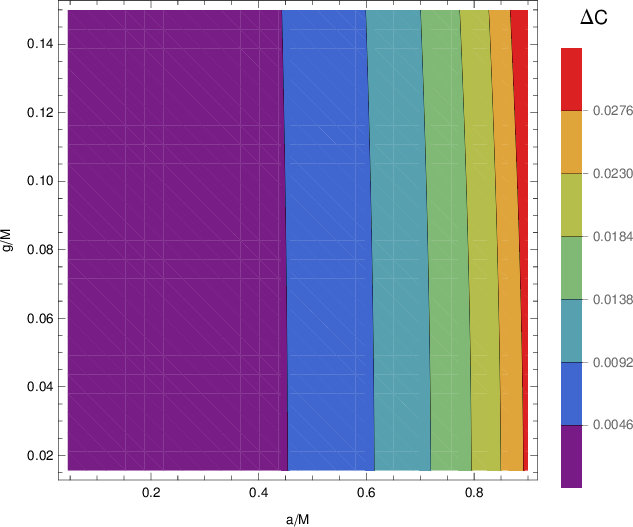}
	\end{minipage}
\caption{The contours corresponding to varying values of \( \Delta c \) are 	depicted in the a vs. g parameter space for fixed parameters \( \omega = 0.8 \) and \( \gamma = 0.25 \) for the incilantion angles $17^0 (left)$ and $ 90^0(right)$.}
\label{Fig9}
\end{figure*}

In Fig. \ref{Fig8} we explore the parameter space of the spin (\(a\)) and inclination angle (\(\theta_0\)) with respect to \(\Delta C\) (left) and \(\delta\) (right). The results indicate that all values within the scanned parameter space satisfy the observational limit of M87\(^*\). Furthermore, Fig. \ref{Fig9} presents the parameter spaces of \(a/M\) vs. \(g/M\) for inclination angles of \(17^\circ\) and \(90^\circ\), respectively. In both cases, the analyzed parameter spaces are consistent with the observational data of M87\(^*\), confirming that the constraints imposed by \(\Delta C\) are satisfied.

The results demonstrate that both the deviation from circularity and fractional deviation parameter values lie within the permissible range for a regular Bardeen black hole in the framework of asymptotically safe gravity (ASG).

These findings reinforce the compatibility of the ASG framework with the observational data of M87\(^*\) and Sgr A$^*$ and establish robust constraints on the coupling parameters and other relevant black hole parameters within this model.

\section{Conclusion}

To explore the effects of quantum gravity, we introduced an improved renormolized, non-singular magnetic monopole and investigated its shadow properties. The distortion of the shadow image of a rotating black hole is an expected consequence of the frame-dragging effect with an increase in spin a. The  effect of running coupling of the quantum corrected gravitational constant has been insignificant in the shadow radius(Fig.1),although the effect of monopole charge seem insignificant in the shadow distortion initally; as the spin parameter increases the distortion s become more pronounced at higher g values.

In addition, we examine the energy conditions associated with the effective stress-energy tensor of the spacetime. We find that the energy density remains positive throughout, thereby satisfying the primary requirement of the weak energy condition. However, the quantity $\mathcal{E} + p_{\theta}$ becomes negative in a localized region near the black hole core, indicating a violation of the null and weak energy conditions induced by quantum corrections. This violation is anisotropic and becomes increasingly concentrated near the equatorial plane with higher spin, while stronger quantum corrections tend to reduce its extent. These features support the interpretation that quantum effects and rotation together contribute to the regular, non-singular nature of the spacetime.

In conclusion, our study provides a detailed characterization of the rotating Bardeen black hole in the background of asymptotically safe gravity (ASG), using observational data from the supermassive black holes $M87^*$ and $Sgr A^*$. By focusing on the coupling parameters $ \gamma $ , $ \omega $ and the magnetic monopole charge g,we see that permisable values of $\Delta C$ are below the 0.1 contour imposing a constraint on $\gamma $ and $g$. Considering the constraint,we demonstrate that the deviation parameter $ \Delta C $ remains within the permissible range for M87$^*$ at all inclination angles. Additionally, we confirm that the parameter spaces for spin ($ a $) and magnetic monopole $( g )$ are consistent with the observational limits for $M87^*$, thus validating the constraints imposed by $\Delta C $.

Further, the observational limits derived from both VLTI and Keck data constrain the fractional deviation parameter $\delta $ to the permissible range of $ -0.14 < \delta < 0.01 $. Our analysis of the parameter space for $ a/M $ and $ \theta_0 $ with respect to $ \delta $ satisfies the permissible range. We can also see that the $\delta$is in the permissible range with  an inclination angle of $ \theta_0 = 50^\circ $ for the parameter space of $ a/M $ vs. $ g/M $ . These results reinforce the compatibility of the ASG framework with observational data from $M87^*$ and $Sgr A^*$, with constraints on the coupling parameters and other relevant black hole parameters. This provides a solid foundation for future studies on the behavior of black holes within the ASG framework and for further testing these models against observational data.

\section{funding}
The author received no specific funding for this work.

\section{Declaration of Interest}
Dr. Sanjit Das and Gowtham Sidharth M declare that they have no known competing financial interests or personal relationships that could have appeared to influence the work reported in the paper.The authors further declare that no financial support, commercial funding, consultancy arrangements, stock ownership, honoraria, paid expert testimony, patents, or other competing professional or personal affiliations have influenced the conception, methodology, analysis, interpretation, or presentation of the results reported in this study. The research was conducted independently and in full compliance with accepted academic and ethical standards. 

\section{Author Contribution}
All the authors contributed equally to this work. 
Gowtham Sidharth M and Sanjit Das jointly carried out the conceptualization, methodology, analytical calculations, preparation of figures, and writing of the manuscript.

\section{data}
No datasets were generated or analysed in this study. All results were derived from the analytical calculations provided in the article.










\end{document}